\documentclass[fleqn,twoside]{article}
\usepackage[headings]{espcrc2A}

\readRCS  
$Id: espcrc2.tex,v 1.2 2004/02/24 11:22:11 spepping Exp $
\usepackage{epsfig}
\usepackage[figuresright]{rotating}

\usepackage{axodraw}
\usepackage{amsmath}
\usepackage{amsfonts}
\usepackage{amsthm}
\usepackage{amssymb}
\usepackage{mathbbol}
\usepackage{array}

\newcommand{\eps}{\varepsilon}

\title{\vspace*{-5mm}
{\footnotesize DESY 06-118 \hfill {\tt 
hep-ph/0607300}}\\
 Evaluating Two-Loop massive Operator Matrix Elements
  with Mellin-Barnes Integrals}

\author{Isabella Bierenbaum\address[MCSD]{Deutsches Elektronen-Synchrotron, DESY, Platanenallee 6, D-15738 Zeuthen, Germany}%
		,
        Johannes Bl\"umlein\addressmark,
        and
        Sebastian Klein\addressmark[MCSD] }

\begin{document}

\begin{abstract}
\noindent We calculate massive 5-propagator 2-loop integrals for operator 
matrix elements in the light-cone expansion, using Mellin-Barnes techniques 
and representations through generalized hypergeometric functions.
\end{abstract}

\maketitle
\pagestyle{empty}
\thispagestyle{empty}

\section{Introduction}

\vspace{1mm}\noindent
The use of Mellin-Barnes integrals became a widespread technique for
calculating Feynman diagrams throughout the last years \cite{MB1}, in particular 
to calculate double and triple box-diagrams. In
Ref.~\cite{TWOLOOP}, it was possible to expand the scalar
two-loop two-point function in all orders in the dimensional
regularization parameter $\varepsilon$, using additionally the gluing
operation of Feynman diagrams, defined by Kreimer \cite{KRE}.  In this
paper, we will apply this technique to a more complex problem, namely
the calculation of massive five--propagator 2--loop Feynman Diagrams with operator
insertions, stemming from light-cone expansion, which are needed for the calculation of the 
heavy flavor coefficient functions in deep--inelastic scattering. These
Wilson coefficients have been calculated before  up to next-to-leading  order
\cite{NLODIS}. Fully analytic results could only be obtained in the limit $Q^2
\gg m^2$ using mass factorization \cite{BUZA}.

The heavy flavor Wilson coefficients are obtained  
as a convolution of the massless Wilson coefficients
$C_{2(L),i}^k\left({Q^2}/{\mu^2},z\right)$ and the
massive operator matrix elements (OME)
${A_{i,j}^{(k)}\left({\mu^2}/{m^2}\right)}$, which shall be calculated
to order $O(\alpha_s^2$). A calculation of these operator matrix elements was
performed in \cite{BUZA} using integration-by-part techniques \cite{Chetyrkin:1981qh}, which
leads to a large number of terms. Expressing the result in terms of 
Nielsen integrals leads to lengthy expressions for most individual 
diagrams and  the complete result.  

This paper offers a different approach, using Mellin-Barnes
integrals, which will enable us to obtain  more simple results, with
the positive side-effect, that we avoid the creation of many of the
above-mentioned canceling intermediate terms. More precisely, we will
directly obtain analytic results in Mellin space, for a general value
of the Mellin variable $N$, using multiple harmonic sums \cite{BK}, which allows 
to compactify the result obtained in $x$--space \cite{BUZA} significantly. 
Before we will do this, we briefly review the
method used in \cite{TWOLOOP} and starting from this proceed to the more
complex cases at hand.  Finally, we will present results obtained in
this way.

\vspace*{-1mm}
\section{The Method}

\vspace{-1mm}\noindent
For the process we want to calculate, we will encounter diagrams with
operator insertions located either at a line or a vertex of the
diagram. In Figure~1 the seven five-propagator integrals contributing
are shown. The powers $\nu_i$ of propagators are one, except for diagram I, 
where
also the power $\nu_1 = 2$ occurs for the line with the operator insertion.
These diagrams will be calculated in a similar way 
to the calculation performed for 
the massless two-loop two-point function, cf. \cite{TWOLOOP,IB}. We start here
with the massive two-loop two-point function, with four massive
propagators (thick lines). The two-loop two-point diagram can be split
into the one-loop two-point and the one-loop three-point function,
using the gluing operation of graphs in $D= 4 - 2 \varepsilon$ dimensions 
\cite{TWOLOOP}:
%
%
\vspace*{-7mm}
\begin{figure}[h]
   \epsfig{file=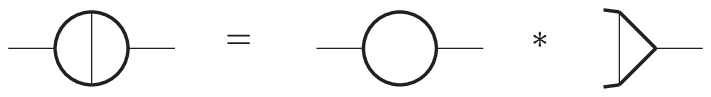,width=6cm}
\end{figure}
%
%
\vspace{-6mm}
\begin{eqnarray}
    =
   \int 
   \frac{d^Dk_1/i\pi^{\frac{D}{2}} \qquad [m^2]^{\nu_{14}-D/2}}
   {[k_1^2+m^2]^{\nu_1}[(k_1-p)^2+m^2]^{\nu_4}}\times 
\nonumber
\end{eqnarray} 
\begin{eqnarray}
   \int  \frac{\hspace{-2cm} d^Dk_2/i\pi^{\frac{D}{2}}\qquad [m^2]^{\nu_{235}-D/2} }
   {[k_2^2+m^2]^{\nu_2}[(k_2-p)^2+m^2]^{\nu_3}[(k_2-k_1)^2]^{\nu_5}}.
\nonumber
\end{eqnarray}
%
%
\begin{figure}[h]
\begin{center}
     \epsfig{file=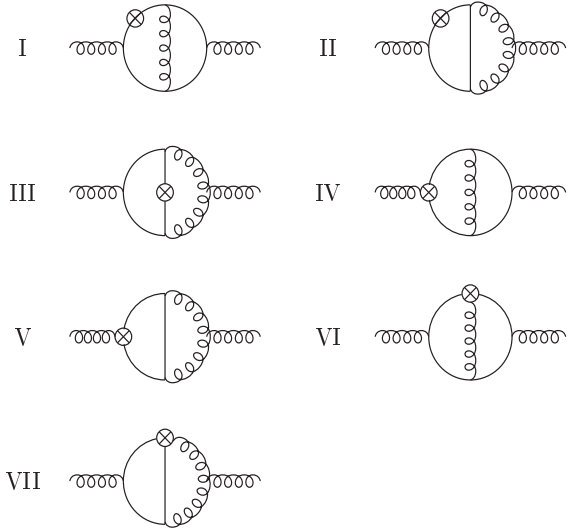,height=7cm,width=7.5cm}
\end{center}
\caption{The seven diagrams with operator insertions; $-\!\!\otimes\!\!- = (\Delta.p)^{N-1}$.}
\label{fig:insertion_rules}
\end{figure}
%
%
\vspace*{-5mm}
  The $*$--operation means that the original 
diagram
  is re-obtained by inserting the three-point function into the
  two-point function.  The same as for the graphs is done on the level
  of integrals, ``splitting'' the two-loop integral into its one-loop
  two-point and three-point components. 
  The insertion of the
  three-point function into the two-point function is reflected by the
  fact that the result of the three-point function alters the result
  of the two-point function by shifting the arguments of its
  propagator-exponents. We start with the three-point
  function:\\ [1em] %
	\SetScale{0.4}%
	\SetWidth{1}%
        \begin{picture}(50,10)(0,0)%
	\Line(70,0)(115,0) %
	\Line(35,35)(35,-35)%
  	\Line(70,0)(35,35) %
	\Line(70,0)(35,-35) %
	\Line(35,35)(20,37)%
  	\Line(35,-35)(20,-37)%
	\end{picture}   %
=
\\[1em]
      \begin{displaymath}
        \int 
        \frac{\hspace{-2cm} d^Dk_2/i\pi^{\frac{D}{2}} \qquad [m^2]^{\nu_{235}-D/2}}
        { [k_2^2+m^2]^{\nu_2}[ (k_2-p)^2+m^2]^{\nu_3}[ (k_2-k_1)^2]^{\nu_5}}
      \end{displaymath}
In a first step, we apply Feynman parameterization to
the propagators, turning the product of three propagators into a sum
of them:
\begin{eqnarray}
    \frac{1}{A_1^{\nu_1}A_2^{\nu_2}...A_n^{\nu_n}} &=&
    \int_0^1 dx_1...dx_n\:
    \delta \left(\sum x_i-1\right) \nonumber\\ 
 &\times& \frac{\Pi x_i^{\nu_i-1}}{\left[\sum x_i A_i\right]^{\sum \nu_i}}
    \frac{\Gamma(\nu_1+...+\nu_n)}{\Gamma(\nu_1)...\Gamma(\nu_n)}. \nonumber
\end{eqnarray}
    We can then group these propagators and apply twice a Mellin-Barnes 
transformation \cite{MELBA}:
\begin{eqnarray}
&&      \frac{1}{(A_1+A_2)^{\nu}}
\nonumber\\
&&      =\frac{1}{2\pi i}\int_{\gamma-i\infty}^{\gamma+i\infty}\hspace{-0.3cm}
      d\sigma \:
      A_1^{\sigma}A_2^{-\nu-\sigma}
      \frac{\Gamma(-\sigma)\Gamma(\nu+\sigma)}{\Gamma(\nu)}.
\nonumber
\end{eqnarray}
  The Mellin-Barnes integral is defined with a contour parallel to the
  imaginary axis, which is placed at a value $\gamma$, such that it
  separates the poles of $\Gamma(-\sigma)$ from the poles of
  $\Gamma(\nu+\sigma)$. Since we have a sum of three propagators, we
  have to apply the Mellin-Barnes integral twice and obtain:
  \begin{eqnarray*} 
   \mbox{I}^{(1,3)} = \frac{c(\Gamma)}{(2\pi i)^2}
  \int_{\gamma_1-i\infty}^{\gamma_1+i\infty} \hspace{-0.3cm} d\sigma
  \int_{\gamma_2-i\infty}^{\gamma_2+i\infty} \hspace{-0.3cm} d\tau
  \: \Gamma(\eps,\nu_i,\sigma,\tau) \\ \times
  \left(\frac{(k_1-p)^2+m^2}{m^2}\right)^{\sigma}
  \left(\frac{k_1^2+m^2}{m^2}\right)^{\tau}. 
  \end{eqnarray*} 
  The term
  $\Gamma(\eps,\nu_i,\sigma,\tau)$ is an abbreviation for a fraction
  of $\Gamma$--functions of the arguments $\eps,\nu_i,\sigma$ and
  $\tau$. The three-point function depends on the
  momenta of the two-point function via the terms with exponents
  $\sigma$ and $\tau$ only. Combining this result with the result of the
  two-point function, which simply is a fraction of two 
  $\Gamma$-functions, shifts their arguments  
by $-\tau$,
  $-\sigma$ respectively: 
  \begin{eqnarray*} \frac
  {\Gamma\,(\nu_{14}-D/2)}{\Gamma\,(\nu_{14})} \longrightarrow \frac
  {\Gamma\,(-\sigma-\tau+\nu_{14}-D/2)}{\Gamma\,(-\sigma-\tau+\nu_{14})}~.
  \end{eqnarray*} 
  In this way, we obtain for the two-loop two-point
  function $\hat{\mbox{I}}^{(2,5)}$: 
\begin{equation}
  \hat{\mbox{I}}^{(2,5)} = \frac{c(\Gamma)}{(2\pi i)^2}
  \int_{\gamma_1-i\infty}^{\gamma_1+i\infty} \hspace{-0.3cm} d\sigma
  \int_{\gamma_2-i\infty}^{\gamma_2+i\infty} \hspace{-0.3cm} d\tau \:
   \: \Gamma'(\eps,\nu_i,\sigma,\tau)~,
\nonumber
\end{equation} 

\begin{table*}[htb]
\caption{The first four Mellin moments for graphs $\rm{I}$ to $\rm{VII}$, 
using M.~Czakon's MB package. All $\nu_i=1$
except for Ib: $\nu_1 = 2$.
}
\label{table:results1}
\newcommand{\m}{\hphantom{$-$}}
\newcommand{\cc}[1]{\multicolumn{1}{c}{#1}}
\renewcommand{\tabcolsep}{0.45pc} 
\renewcommand{\arraystretch}{1.2} 
\begin{tabular}{@{}lllll}
\hline\hline
$\displaystyle {N}$  & \cc{$\displaystyle 2$} & \cc{$\displaystyle 3$} & \cc{$\displaystyle 4$} 
& \cc{$\displaystyle 5$} \\
\hline\hline
$\displaystyle \rm{Ia}$  & \m $\displaystyle +0.49999$ & \m $\displaystyle +0.31018$ 
& \m $\displaystyle +0.21527$ & \m $\displaystyle +0.16007$ \\
\hline
$\displaystyle \rm{Ib}$  
&\m $\displaystyle -0.09028$
&\m $\displaystyle -0.04398$
&\m $\displaystyle -0.02519$ 
&\m $\displaystyle -0.01596$
\\
\hline
$\displaystyle \rm{II}$  
& \m $\displaystyle -0.24999~\eps^{-1}$ 
& \m $\displaystyle -0.15277~\eps^{-1}$ 
& \m $\displaystyle -0.10416~\eps^{-1}$ 
& \m $\displaystyle -0.07611~\eps^{-1}$ \\
& \m $\displaystyle +0.53861$ 
& \m $\displaystyle +0.33609$ 
& \m $\displaystyle +0.23483$ 
& \m $\displaystyle +0.17573$ \\
\hline
$\displaystyle \rm{III}$  
& \m $\displaystyle O(10^{-17})~\eps^{-1}$ 
& \m $\displaystyle -0.04166~\eps^{-1}$ 
& \m $\displaystyle O(10^{-16})~\eps^{-1}$ 
& \m $\displaystyle -0.01111~\eps^{-1}$ \\
& \m $\displaystyle O(10^{-6})$ 
& \m $\displaystyle +0.06893$ 
& \m $\displaystyle O(10^{-6})$ 
& \m $\displaystyle+0.016527$ \\
\hline
$\displaystyle \rm{IV}$  & \m $\displaystyle +0.99999$ & \m $\displaystyle 0.$ & \m $\displaystyle +0.43055$ & \m 
$\displaystyle O(10^{-6})$ \\
\hline
$\displaystyle \rm{V}$  
& \m $\displaystyle -0.49999~\eps^{-1}$ 
& \m $\displaystyle O(10^{-17})~\eps^{-1}$ 
& \m $\displaystyle -0.20833~\eps^{-1}$ 
& \m $\displaystyle O(10^{-17})~\eps^{-1}$ \\
& \m $\displaystyle +1.07722$ 
& \m $\displaystyle +O(10^{-12})$ 
& \m $\displaystyle +0.46967$ 
& \m $\displaystyle +O(10^{-9})$ \\
\hline
$\displaystyle \rm{VI}$  & \m $\displaystyle +0.99999$ 
& \m $\displaystyle +0.99999$ & \m $\displaystyle +0.90277$ & \m 
$\displaystyle +0.80555$ \\
\hline
$\displaystyle \rm{VII}$  
& \m $\displaystyle -0.49999~\eps^{-1}$ 
& \m $\displaystyle -0.24999~\eps^{-1}$ 
& \m $\displaystyle -0.20833~\eps^{-1}$ 
& \m $\displaystyle -0.13888~\eps^{-1}$ \\
& \m $\displaystyle +1.07723$ 
& \m $\displaystyle +0.53862$ 
& \m $\displaystyle +0.44189$ 
& \m $\displaystyle +0.30616$ \\
\hline\hline
\end{tabular}\\[2pt]
\end{table*}

\noindent
  where $\Gamma'(\eps,\nu_i,\sigma,\tau)$ now consists
  of $\Gamma$-functions stemming from the result of the two-point and the
  three-point function. $c(\Gamma)$ is a factor, which contains
  constants and other $\Gamma$-functions not depending on the
  arguments $\sigma$ and $\tau$. Closing the contour at infinity and collecting the
  residues of the $\Gamma$-functions inside the integration area via:
  $\mbox{res}\,\big[\Gamma(-x+a),\;x=a+n\big]={\displaystyle
  -\frac{(-1)^n}{n!}}$, we are left with a result which consists in 
  general
  of (infinite) sums of $\Gamma$-functions:
\begin{eqnarray} 
 \hat{\mbox{I}}^{(2,5)}&=&
  c(\Gamma)\:
  \sum_{n=0}^{\infty}\sum_{j=0}^{\infty}
\frac{\Gamma(n\pm b_1)\Gamma(n\pm b_2)}{n!\; \Gamma(n\pm b_3)}\nonumber\\
& &  \phantom{\sum_{n=0}^{\infty}\sum_{j=0}^{\infty}}\times 
\frac{\Gamma(j\pm c_1)\Gamma(j\pm c_2)}{j!\; \Gamma(j\pm c_3)}\nonumber\\
& &  \phantom{\sum_{n=0}^{\infty}\sum_{j=0}^{\infty}}\times 
\frac{\Gamma(n+j\pm a_1)\Gamma(n+j\pm a_2)}{\Gamma(n+j\pm a_3)\Gamma(n+j\pm a_4)}.
\nonumber 
\end{eqnarray}
These sums can be calculated with 
the computer libraries {\tt nestedsums} \cite{nestedsums} 
{\tt (C++)} or {\tt XSummer} \cite{XSUMMER}
{\tt (FORM)}.

\section{Operator Matrix Elements}

\vspace{1mm}\noindent
We now turn to the calculation of the diagrams with operator insertion.
The previous calculation is changed due to the emergence of  an additional numerator structure. As an
example, let us consider graph  I of
Figure \ref{fig:insertion_rules}:\\[1em]
  \begin{minipage}[c]{5cm}
   \hspace{0.6cm}
    \epsfig{file=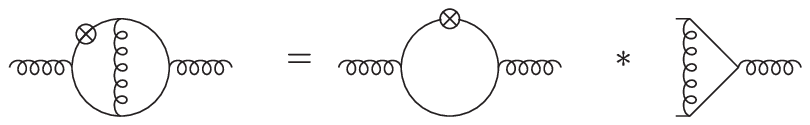,width=6cm}
  \end{minipage}\\[0.5em]
  The two-point function now carries an operator insertion, while the
  three-point function is unaltered. To check for the change of the
  two-point function by the insertion, we first apply again the Feynman
  parameterization to it:
%
%
{\small
\begin{table*}[htb]
\caption{The first four Mellin moments for graphs $\rm{I}$ to $\rm{VII}$. 
$\nu_i=1$; Ib: $\nu_1 = 2$.}
\label{table:results2}
\newcommand{\m}{\hphantom{ }}
\newcommand{\cc}[1]{\multicolumn{1}{c}{#1}}
\renewcommand{\arraystretch}{2.0} 
\begin{tabular}{lcccc}
\hline\hline
$\displaystyle \rm{N}$  & \cc{$\displaystyle 2$} & \cc{$\displaystyle 3$} & \cc{$\displaystyle 4$} & 
\cc{$\displaystyle 5$} \\
\hline\hline
$\displaystyle \rm{Ia}$  & \m $\displaystyle \frac{1}{2}$ & \m $\displaystyle \frac{67}{216}$ 
& \m $\displaystyle \frac{31}{144}$ & \m $\displaystyle \frac{2161}{13500}$ \\
$\displaystyle \rm{Ib}$  
& \m $\displaystyle -\frac{13}{144}$ 
& \m $\displaystyle -\frac{19}{432}$ 
& \m $\displaystyle -\frac{17}{675}$ 
& \m $\displaystyle -\frac{431}{27000}$ \\
$\displaystyle \rm{II}$  & \m $\displaystyle -\frac{1}{4\eps}+\frac{1}{4}+\frac{1}{2}\gamma_E$ 
& \m $\displaystyle -\frac{11}{72\eps}+\frac{23}{144}+\frac{11}{36}\gamma_E$ 
& \m $\displaystyle -\frac{5}{48\eps}+\frac{11}{96}+\frac{5}{24}\gamma_E$ & \m $\displaystyle -\frac{137}{1800\eps}+\frac{949}{10800}+\frac{137}{900}\gamma_E$ \\
$\displaystyle \rm{III}$  
& \m $\displaystyle 0$ 
& \m $\displaystyle -\frac{1}{24\eps}+\frac{1}{48}+\frac{1}{12}\gamma_E$ 
& \m $\displaystyle 0$ 
& \m $\displaystyle -\frac{1}{90}+\frac{1}{270}+\frac{1}{45}\gamma_E$ \\
$\displaystyle \rm{IV}$  & \m $\displaystyle 1$ & \m $\displaystyle 0$ & \m $\displaystyle \frac{31}{72}$ & \m $\displaystyle 0$ \\
$\displaystyle \rm{V}$  & \m $\displaystyle -\frac{1}{2\eps}+\frac{1}{2}+\gamma_E$ & \m $\displaystyle 0$ & \m $\displaystyle -\frac{5}{24\eps}+\frac{11}{48}+\frac{5}{12}\gamma_E$ & \m $\displaystyle 0$ \\
$\displaystyle \rm{VI}$  & \m $\displaystyle 1$ 
& \m $\displaystyle 1$ 
& \m $\displaystyle \frac{65}{72}$ 
& \m $\displaystyle \frac{29}{36}$ \\
$\displaystyle \rm{VII}$  
& \m $\displaystyle -\frac{1}  {2\eps}+\frac {1}  {2}+\gamma_E$ 
& \m $\displaystyle -\frac{1}  {4\eps}+\frac {1}  {4}+\frac{\gamma_E}{2}$
& \m $\displaystyle -\frac{5}{24\eps}+\frac{29}{144}+\frac{5}{12} \gamma_E$
& \m $\displaystyle -\frac{5}{36\eps}+\frac {7} {48}+\frac{5}{18} \gamma_E$\\
[3mm]
\hline\hline
\end{tabular}\\[2pt]
\end{table*}
\renewcommand{\arraystretch}{1.0} 
}
\begin{equation}
\begin{split}
&\hat{\mbox{I}}^{(1,2)}=
\frac{\Gamma(\nu_{14})}{\Gamma(\nu_1)\Gamma(\nu_4)}
(m^2)^{\nu_{14}-D/2}\times
\\
&\int_0^1dx_1dx_2 \; x_1^{\nu_1-1}x_2^{\nu_4-1}
\delta \left(x_1+x_2-1\right)  \times
\\ 
&\int \!\!\frac{d^Dk_1}{i\pi^{\frac{D}{2}}}
\frac{(\Delta.k_1)^{N-1}}
{\left(x_1k_1^2+x_1m^2+x_2(k_1-p)^2+x_2m^2\right)^{\nu_{14}}}.
\end{split}
\nonumber
\end{equation}
Throughout the calculation, we have to shift $k_1 \rightarrow
k_1+x_2p$, leading to a corresponding shift in the numerator $(\Delta.k_1)^{N} \rightarrow 
(\Delta.(k_1+x_2p))^{N}$, which can be expressed via a binomial sum to be:
\begin{displaymath}
  (\Delta.k_1 + x_2 \Delta.p)^{N} =
  \sum_{l=0}^{N} \binom{N}{l} (\Delta.k_1)^l 
  (x_2 \,  \Delta.p)^{N-l}
\end{displaymath}

\noindent
Since $\Delta$ is a light-cone vector and hence $\Delta^2=0$, we
find that all integrals with $(\Delta.k_1)^l$ vanish, except the
one for $l=0$. This leads  to a result for the two-point
function, which is very similar to the original one, but which now
contains the Mellin parameter $N$:\\
%
\begin{equation}
\hat{\mbox{I}}^{(1,2)}=
(\Delta.p)^{N-1}
\frac{\Gamma(\nu_{14}-D/2)\Gamma(\nu_4+N-1)}{\Gamma(\nu_4)\Gamma(\nu_{14}+N-1)}~.
\nonumber
\end{equation}
Here, we use $\nu_{14} \equiv  \nu_1+\nu_4$, etc.
Inserting the three-point function, we obtain for this graph  I:
%
{
\begin{eqnarray}
\hat{I}_{G1} &=&
\frac{1}{(2\pi i)^2}
  \frac{(\Delta.p)^{N-1}}{\Gamma(\nu_2)\Gamma(\nu_3)\Gamma(\nu_5)\Gamma(D-\nu_{235})}\nonumber\\
  &&\: \times\int_{\gamma_1-i\infty}^{\gamma_1+i\infty} \hspace{-0.3cm} d\sigma
  \int_{\gamma_2-i\infty}^{\gamma_2+i\infty} \hspace{-0.3cm} d\tau \:
  \Gamma(-\sigma)\Gamma(\nu_3+\sigma)\nonumber\\
   &&\: \times \frac{\Gamma(-\sigma+\nu_4+N-1)}{\Gamma(-\sigma+\nu_4)}\Gamma(-\tau)
\Gamma(\nu_2+\tau)\nonumber\\
  &&\: \times \frac{\Gamma(\sigma+\tau+\nu_{235}-D/2)\Gamma(\sigma+\tau+\nu_5)}
{\Gamma(\sigma+\tau+\nu_{23})}\nonumber\\
  &&\: \times \Gamma(-\sigma-\tau+D-\nu_{23}-2\nu_5)\nonumber\\
&&\: \times \frac{\Gamma(-\sigma-\tau+\nu_{14}-D/2)}{\Gamma(-\sigma-\tau+\nu_{14}+N-1)}.\nonumber
\nonumber
\end{eqnarray}
}

\begin{table*}[htb]
\caption{The analytic results for graphs I to VII for general values of N, with all $\nu_i = 1$, Ib: $\nu_1 = 2$.}
\label{table:results3}
\newcommand{\m}{\hphantom{$-$}}
\newcommand{\cc}[1]{\multicolumn{1}{c}{#1}}
\renewcommand{\arraystretch}{2.3} 
\begin{tabular}{ll}
\hline\hline
$\rm{Ia}$            & \m $\displaystyle 
\frac{{S}_1^2({N})+3{S}_2({N})}{2{N}({N}+1)}$ \\
$\rm{Ib}$            & \m $\displaystyle 
\frac{{S}_1({N})-S_2(N)-S_{1,1}(N)}{{N}({N}+1)(N+2)}- \frac{1}{(N+1)^2(N+2)}$ \\
$\rm{II}$           & \m $\displaystyle 
\frac{{S}_1({N})}{{N}({N}+1)}\left(-\frac{1}{\eps} + 2 \gamma_E\right)
+2\frac{{S}_1({N})}{({N})({N}+1)^2}+\frac{{S}_1^2({N})-{S}_2({N})}{2{N}({N}+1)}$  
\\
$\rm{III}$          & \m $\displaystyle 
\frac{[1-(-1)^{{N}}]}{{N}({N}+1)^2}\left(-\frac{1}{\eps}+\frac{2}{({N}+1)}+2\gamma_E\right)$  \\
$\rm{IV}$           & \m $\displaystyle [1+(-1)^{{N}}] \times \rm{I_a}$  \\
$\rm{V}$            & \m $\displaystyle [1+(-1)^{{N}}] \times \rm{II}$  \\
$\rm{VI}$            & \m $\displaystyle \frac{4}{N} \left[S_2(N) - \frac{S_1(N)}{N} \right]$ \\ 
$\rm{VII}$            &  
\m $\displaystyle 
\left[\frac{(-1)^N -1}{N^2(N+1)} + \frac{2S_1(N)}{N(N+1)}\right]\left(-\frac{1}{\eps} + 2 
\gamma_E\right)$
\\ & \m $\displaystyle
+\left[2\frac{(-1)^N -1}{N^2(N+1)^2} + 
\frac{S_1^2(N)-S_2(N)+2S_{-2}(N)}{N(N+1)}+\frac{2(3N+1)S_1(N)}{N^2(N+1)^2}\right]$\\
[3mm]
\hline 
\hline
\end{tabular}\\[2pt]
\end{table*}

\noindent
Analogously, we built the Mellin-Barnes integrals for the
remaining six graphs and used the {\tt mathematica} package {\tt MB} by M.~Czakon
\cite{MB}, to numerically produce the results for the first few Mellin
moments, given in Table~1. They serve as a check for our analytic result.  

To derive the analytic results, we continued from here and used relations like
\begin{equation}
\begin{split}
& \frac{1}{2\pi i}\int_{-\infty}^{+\infty}\: ds
 \Gamma(a+s)\Gamma(b+s)\Gamma(d-a-b-s)
\\
& \hspace{2cm}\times \: \frac{\Gamma(e-c+s)\Gamma(-s)}{\Gamma(e+s)}
\\
& =\frac{\Gamma(e-c)\Gamma(a)\Gamma(b)\Gamma(d-a)\Gamma(d-b)}{\Gamma(e)\Gamma(d)}
\\
& \hspace{2cm} \times \:\: \empty{}_3F_2[a,b,c;d,e;1] \: ,
\end{split}\nonumber
\end{equation}
and other relations which are more general than Barnes formulae \cite{SLAT}. 
For the second
integration we usually applied the Residue Theorem and in this way
obtained double sums that contain the symbolic parameter $N$:
{\small 
\begin{equation}
\begin{split}
&\hat{I}_{G1}
\Rightarrow
(\Delta.p)^{N-1}\:
\frac{\Gamma(N+1)}{\Gamma(1-2\eps)}
\:\sum_{k=0}^{\infty} \sum_{j=0}^{\infty} 
\frac{\Gamma(k+1)}{\Gamma(k+2+N)} 
\end{split}\nonumber \end{equation} 
\begin{equation} \begin{split}
&\!\! \times \Bigl[ 
\Gamma(\eps)\Gamma(1-\eps)\times
\\
&\!\! \frac{\Gamma(j+1-2\eps)\Gamma(j+1+\eps)}{\Gamma(j+1-\eps)\Gamma(j+2+N)} 
\frac{\Gamma(k+j+1+N)}{\Gamma(k+j+2)} 
\\
&\!\! +
\Gamma(-\eps)\Gamma(1+\eps)\times
\\
&\!\!
\frac{\Gamma(j+1+2\eps)\Gamma(j+1-\eps)}{\Gamma(j+1)\Gamma(j+2+\eps+N)} 
\frac{\Gamma(k+j+1+\eps+N)}{\Gamma(k+j+2+\eps)} 
\Bigr].
\end{split}\nonumber
\end{equation}
}
\noindent
Sums like these containing a symbolic parameter $N$ cannot be done
via {\tt nestedsums} or {\tt Xsummer}, except 
for the most simple cases, fixing $N$. 
For other more complicated cases, however, we had to use 
special transformations.
As a computer algebra system we used {\tt MAPLE}. For the first four Mellin moments 
the results are shown in
Table~2 and agree with the numeric results obtained by
the use of {\tt MB}. 
The graphs IV and V are, for all values of $\nu_i = 1$, related to the graph Ia, respectively
II, by a simple factor $[1+(-1)^{N}]$ due to the fact that the operator insertion is located
on a three-vertex with an on-shell external line with $p^2 = 0$.
Diagrams~VI and VII require special treatment. The corresponding integrals
are evaluated most effectively first transforming them into generalized hypergeometric 
functions \cite{SLAT}. For fixed values of $N$ again these analytic results agree
with the numerical values presented in Table~1.

Finally, we derive the complete analytic result for general values of $N$. To obtain
this, we had to extensively use algebraic and analytic relations
to convert the sums obtained in evaluating the Mellin--Barnes integrals, 
into expressions containing 
(nested) harmonic sums and related objects in intermediary steps.
The final results depend on harmonic sums only and are summarized in Table~3. 
From this form the analytic continuation to complex values of $N$ \cite{BK,ANCONT1,BM1}
needed in data analyzes~\footnote{A fast numerical analytic continuation of the heavy 
flavor Wilson coefficients \cite{NLODIS} was given in \cite{AB}.} can be performed directly.
Although the diagrams I--VII are the most sophisticated in the 2--loop problem under consideration,
the results turn out to be very simple and much of the complexity of the Wilson coefficient stems from other,
more simple structures.

In summary, we have shown that  by use of Mellin--Barnes integrals and direct representations
through generalized hypergeometric functions, massive two--loop five--propagator integrals containing
operator insertions can be calculated in a very effective way in terms of nested harmonic sums.
The corresponding expressions have a rather simple structure and can be expressed even in terms
of only single harmonic sums. The integration-by-part method, on the other hand, allows to treat
diagrams of lower complexity (4-propagator integrals), but leads to results with a much more complex 
structure.


\noindent
{\bf Acknowledgments:}~This work was supported in part by DFG
Sonderforschungsbereich Transregio 9, Computergest\"utzte Theoretische
Physik, [SFB/CPP-06-35].


\end{document}